%% file: REPAC_ICASSP21_main.tex
\documentclass{article}
\usepackage{spconf,amsmath,graphicx}
\usepackage[sorting=none]{}

\usepackage[absolute]{textpos}
\usepackage{everyshi}

\title{REPAC: RELIABLE ESTIMATION OF \\ PHASE-AMPLITUDE COUPLING IN BRAIN NETWORKS}

\name{Giulia Cisotto\thanks{This work was supported by REPAC project funded by the University of Padova under the initiative SID-Networking 2019. Part of this work was also supported by MUR (Ministry of University and Research) under the initiative Departments of Excellence (Law 232/2016).}}
\address{
\textit{Dept. Information Engineering, University of Padova}, Padova, Italy \\
\textit{National Centre for Neurology and Psychiatry}, Tokyo, Japan \\
\textit{National Inter-University Consortium for Telecommunications (CNIT)}, Padova, Italy \\
}

\usepackage[acronyms,nonumberlist,nopostdot,nomain,nogroupskip]{glossaries}
\input{./acronyms.tex}

\begin{document}

\maketitle

\input{./Abstract.tex}

\begin{textblock*}{17cm}(1.7cm, 0.5cm)
	\noindent\scriptsize This work has been submitted to the IEEE for possible publication. Copyright may be transferred without notice, after which this version may no longer be accessible.\\
	\textbf{Copyright Notice}: \textcopyright 2020 IEEE. Personal use of this material is permitted. Permission from IEEE must be obtained for all other uses, in any current or future media, including reprinting/republishing this material for advertising or promotional purposes, creating new collective works, for resale or redistribution to servers or lists, or reuse of any copyrighted component of this work in other works.
\end{textblock*} 

\begin{keywords}
phase-amplitude coupling, brain networks, REPAC, modulation, EEG
\end{keywords}

\section{Introduction}\label{sec:Intro}
\input{./Intro.tex}

\section{Methods}\label{sec:Methods}
\input{./Methods.tex}

\section{Results}\label{sec:Results}
\input{./Results.tex}

\section{Discussion}\label{sec:Discussion}
\input{./Discussion.tex}

\section{Conclusions and future works}

PAC mechanisms are recently getting increasing attention, given their potential role in shedding light on the entanglement and disentanglement of important brain networks, e.g., the fronto-parietal network, enabler of the working memory and other sensory functions.
This contribution presents REPAC, a reliable and robust algorithm to identify and characterize PAC events. 
Some limitations still affect this work and will deserve further investigations. 
For example, making \gls{repac} decode the HFO bursts embedded into the LFO \emph{peaks} should be included, as
some recent works explained this coupling mechanism with the functioning of the fronto-posterior network during working memory tasks~\cite{Sauseng2016}. 
Moreover, phase precession of LFO can be considered in the synthetic PAC model~\cite{Lisman2013}. 
Finally, investigating non-sinusoidal waveforms driving the PAC could help understanding how the brain accomplishes other complex tasks \cite{Cole2017}. 
Then, a more extensive campaign of tests on simulated and real signals should be performed to further validate REPAC.
%
%
Given its increased detection performance, \gls{repac} paves the road to a more efficient modeling of the PAC phenomenon that could provide new significant insights for the explanation of brain mechanisms in many different conditions~\cite{Silvoni2014, AESM2018, HIT2020}, both healthy and pathological.

%


\vfill\pagebreak


\bibliography{References_new.bib}
\bibliographystyle{IEEEbib}

\end{document}

%% file: acronyms.tex
\newacronym{pac}{PAC}{phase-amplitude coupling}
\newacronym{repac}{REPAC}{robust estimation of phase-amplitude coupling}
\newacronym{mvl}{MVL}{mean vector length}

%% file: Abstract.tex
\begin{abstract}

Recent evidence has revealed cross-frequency coupling and, particularly, phase-amplitude coupling (PAC) as an important strategy for the brain to accomplish a variety of high-level cognitive and sensory functions. However, decoding PAC is still challenging. This contribution presents REPAC, a reliable and robust algorithm for modeling and detecting PAC events in EEG signals.
First, we explain the synthesis of PAC-like EEG signals, with special attention to the most critical parameters that characterize PAC, i.e., SNR, modulation index, duration of coupling. Second, REPAC is introduced in detail.
We use computer simulations to generate a set of random PAC-like EEG signals
and test the performance of REPAC with regard to a baseline method.
REPAC is shown to outperform the baseline method
even with realistic values of SNR, e.g., $-10$~dB.
They both reach accuracy levels around $99\%$, but REPAC leads to a significant improvement of sensitivity, from $20.11\%$ to $65.21\%$, with comparable specificity (around $99\%$).
REPAC is also applied to a real EEG signal showing preliminary encouraging results.








\end{abstract}

%% file: Intro.tex

Cross-frequency coupling (CFC) generally labels the interaction between two components at different frequencies in the brain~\cite{Canolty2010}. 
Among the most common CFC mechanisms, the synchronization between the amplitude of \emph{high frequency oscillations} (HFO) and the phase of \emph{low frequency oscillations} (LFO), called \emph{phase-amplitude coupling (PAC)}, has been found in many electrophysiological experiments, both in humans and in animals, using invasive as well as non-invasive signal acquisition methods~\cite{Fontolan2014,Healthcom2018}. 
Depending on the function to be accomplished, coupling can occur at the troughs~\cite{Canolty2010, Tort2010, DeHemptinne2013} or at the peaks~\cite{Fell2011} of the slower component, i.e., the lower frequency band. 
However, in most cases, HFO bursts (i.e., shortly lasting) have been associated with the troughs of the LFO band, when sensory processing, decision-making and other cognitive functions are accomplished~\cite{DeHemptinne2013}. 
Several metrics have been already implemented to estimate the strength of such interaction, i.e., the phase-locking value (PLV)~\cite{Lachaux1995}, the mean vector length (MVL)~\cite{Canolty2010} and the Kullback-Leibler modulation index (MI)~\cite{Tort2010}.
However, they typically choose a too narrow LFO band, excluding potentially PAC-related information from the estimation, and the HFO band is generally either defined a-priori, based on previous empirical field knowledge, or kept as larger as possible.
%
%
This approach might have led to that variability of results in the current literature that makes it difficult to compare different studies, even when they use the same PAC metric.
Therefore, this contribution aims (i) to identify the main parameters involved in the coupling and provide a way to generate PAC-like synthetic EEG signals, (ii) to present a \gls{repac}, with a novel fully data-driven model of the high frequency component, and (iii) to provide a discussion about its performance, as well as its current limitations.


%% file: Methods.tex

This section introduces the synthetic model to simulate EEG signals including some PAC events. Then, the newly proposed REPAC algorithm is presented.

\subsection{Synthesis of a typical PAC-like EEG signal}\label{sec:synthesis}
\begin{figure*}[htpb]
    \begin{center}
	\includegraphics[width=0.45\textwidth, height=0.3\textwidth]{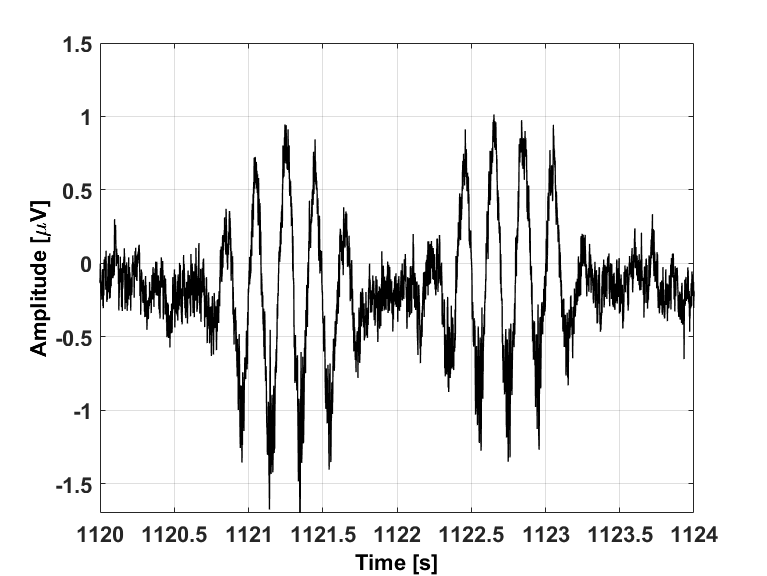}(a)
    \includegraphics[width=0.45\textwidth, height=0.3\textwidth]{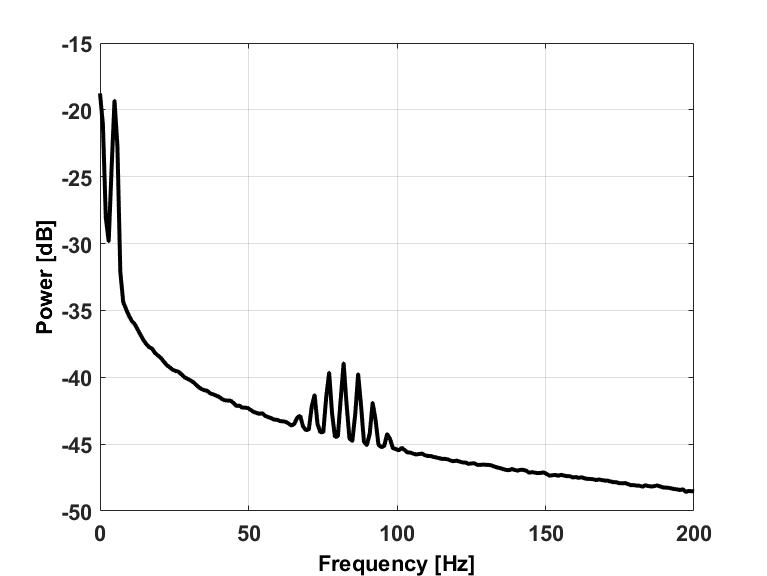}(b)
	\caption{An example of $4$~s PAC-like signal with $f_L = 5$~Hz, $f_H = 80$~Hz, $m = 0.1$, $L = 1.5$~s and $SNR = -5$~dB. (a) Time course and (b) Power spectrum.}
    \label{fig:SYNTHSIGN}
    \end{center}
\end{figure*}

As for the synthesis of a typical PAC embedded EEG signal, we extend the model proposed by Miyakoshi and collaborators in \cite{Makoto2013}: a random number of PAC events are added to a pink noise signal to achieve a specific signal-to-noise-ratio ($SNR$).
For the sake of simplicity, all frequency components involved in the PAC are considered as pure sinusoids, i.e., oscillations. 
A few cycles of a lower frequency sinusoid (usually in the order of a few Hz) are typically considered to simulate a PAC event; then, at each of its trough, a burst of the higher frequency sinusoid (usually above 30 Hz) is embedded. This means that the highest frequency sinusoid oscillates during the duration of the trough and its amplitude is modulated by the amplitude of the lower frequency sinusoid \cite{Makoto2013}.
As additional novelty compared to the current literature, we assume that the LFO - itself - is (very) slowly modulated in amplitude. This assumption is necessary in order to have (periodic) PAC events in the signal, on the top of other brain activities in the same LFO band.
This results in a typical PAC-like signal, as shown in Fig.~\ref{fig:SYNTHSIGN}. To note, its power spectral density (PSD) displays three main components: two in the low frequency range and one (more complex) in the high frequency range. The former reflect the presence of a low frequency component that is modulated in amplitude during the course of the time; the latter, instead, consists in a family of peaks around the central frequency of the high frequency component.
%
%
%
%
Then, in the construction of the signal model, three main parameters are considered: the $SNR$, the modulation index ($m$) and the length of the PAC event ($L$).
Since non-invasive EEG recordings are typically characterized by very low SNRs \cite{Ball2009, Globecom2018}, in this contribution we consider the values of ${-18, -10, -5, 0}$~dB for simulations. 
On the other hand, $m$ accounts for the strength of the coupling between the low and the high frequency component. It takes values in the range $[0, 1]$: the closer to 1, the higher the correlation between the amplitude of the low frequency sinusoid and that of the high frequency one. 
Finally, the length of the PAC event, namely $L$, has been rarely taken into account: here, values between $1.5$~s and $5$~s are used. Nevertheless, this parameter could strongly depend on the specific application or pathology under investigation. 


\subsection{REPAC: robust estimation of PAC}
Let $s[n]$ be a synthetic EEG signal, built as in Section~\ref{sec:synthesis} and including some PAC events (see Fig.~\ref{fig:SYNTHSIGN}).
First, the trial-and-error procedure, as implemented in PACT \cite{Makoto2013} (a Matlab-based toolbox included in EEGLAB \cite{EEGLAB}), is used to preliminarily define an LFO frequency band for the estimation of the coupling.
Also, a candidate HFO band is fixed, a-priori.
Here, however, a novel procedure is proposed to refine the selection of the LFO band: indeed, to ensure that the full LFO band involved in the PAC mechanism is considered, a larger LFO is identified as follows.
The raw signal $s[n]$ is filtered in $K$ different low frequency narrow bands to get the low frequency signals $\{s_{LFO}[n]\}_k$ with $k=1,2,...,K$.
Separately, $s[n]$ is also filtered in HFO to get the high frequency signal $s_{HFO}[n]$.
Then, the Hilbert transform \cite{Mitra} is used to extract the phase signals $\{\phi_L[n]\}_k$, with $k=1,2,...,K$, from each $\{s_{LFO}[n]\}_k$ and the envelope signal $A_H[n]$ from $s_{HFO}[n]$.
%
Finally, at each time instant, $n$, and for each $k$, we form a vector with length equal to $A_H[n]$ and with an angle equal to $\phi^{k}_L[n]$.
Then, the \gls{mvl} is computed as the average through time (i.e., time instants $n$)\cite{Tort2010}.
The \gls{mvl} can be interpreted as the average amount of dependence between the phase and the amplitude signals, i.e., in other words their coupling strength.
%
%
Importantly, it has to be noted that not all samples of $A_H[n]$ are taken into account to compute MVL, but only a fraction of them, given by a percentage value named $has$ (to be in line with PACT terminology).
Therefore, for each selection of $has$ and $k$, we obtain an \gls{mvl} value. We average across different $has$ values and we finally get $k$ \gls{mvl} values.
Then, the minimum and the maximum \gls{mvl} values ($MVL_{min}$ and $MVL_{max}$) are used to compute the difference:
$$\Delta_{MVL} = MVL_{max} - MVL_{min}$$
and a threshold value $MVL_{th}$ is defined as $MVL_{max} - 0.1*\Delta_{MVL}.$
Incidentally, the coefficient $0.1$ has been empirically selected. 
The frequency range where the averaged $MVL$ takes values above $ MVL_{th}$ is selected as \emph{refined} LFO band. 
To note, the latter could possibly include more than one narrow band (identified by $k$).
$LFO_{bw}$ represents the bandwidth of such frequency range.
Next, $s[n]$ is filtered in the refined LFO band providing an estimate of the low frequency signal, $\hat{s}_{LFO}[n].$
Given that the phase of a sinusoid linearly increases with the frequency, i.e., $\psi[n] = 2\pi fn$ (in the ideal case), and that the LFO component is assumed to be modulated in amplitude, from the empirical measure of its phase, namely $\hat{\phi}_L[n]$, we can get an estimate of the main frequency $\hat{f}_L$.
Therefore, the (unwrapped) phase of $\hat{s}_{LFO}[n]$ is taken by means of the Hilbert transform and its slope $m_{\hat{\phi}}$ is extracted providing:
$\hat{f}_L = \frac{m_{\hat{\phi}}}{2\pi}.$
Finally, according to the well-known AM theory \cite{Schwartz}, $\hat{s}_{LFO}[n]$ is demodulated to find its slowly \emph{modulating signal} $\hat{s}_1[n]$: specifically, the instantaneous power of $\hat{s}_{LFO}[n]$ is computed and an ideal low pass filter, 
with cut-off frequency equal to 2 Hz 
is applied~\cite{ICC2013}, giving the signal $\hat{s}_1[n]$ (since $\hat{s}_1[n]$ represents the instantaneous power, it is always zero or positive valued).
Assuming that PAC events could be reasonably found during intervals of time where the LFO component is at its largest negative values (i.e., the troughs)~\cite{Fontolan2014, Sauseng2016}, the candidate PAC periods are identified as the intervals of time where the signal $\hat{s}_1[n]$ takes positive (non-zero) values. 
Thus, a number $N$ of PAC periods are identified and the raw signal $s[n]$ is segmented into $N$ segments, containing one PAC period each.
%
In order to extract the HFO component, the power spectrum from every segment is computed and averaged to provide a data-driven estimate of the power spectrum of the HFO component (similar to Fig.~\ref{fig:SYNTHSIGN}b). 
As expected, 
the average power spectrum is a \emph{frequency comb}~\cite{FemtoSecondLASERS} with a central peak at $\hat{f}_H$ and some side peaks, equally-spaced by $\hat{f}_L$. The number of side peaks with non-negligible power is roughly $4$.
Therefore, $HFO_{bw}$ is set to $8*\hat{f}_L$ and the \emph{refined} HFO band is defined as: $(\hat{f}_H - 4 \hat{f}_L$, $\hat{f}_H + 4 \hat{f}_L)$.
The raw signal $s[n]$ is thus filtered in this new HFO band, providing an estimate of the high frequency PAC component, $\hat{s}_{HFO}[n]$.
In order to get a better estimation of $f_H$, the (unwrapped) phase $\hat{\phi}_H[n]$ of $\hat{s}_{HFO}[n]$ is taken by means of the Hilbert transform. Finally, the slope of $\hat{\phi}_H[n]$ is computed and used as estimate of $\hat{f}_H$.
The Hilbert transform is also used to extract the amplitude $\hat{A}_H[n]$ from $\hat{s}_{HFO}[n]$.
Using $\hat{A}_H[n]$ and $\hat{\phi}_L[n]$, a vector is formed at each time point $n$, as explained above, and the final \gls{mvl} is computed as the average along time.

%% file: Results.tex
%
%
This section provides a comparison between \gls{repac} and the baseline method (i.e., PACT \cite{Makoto2013}), based on their ability to identify and characterize PAC events.
%
%
%
%
\begin{figure}[htbp]
\begin{center}
	\includegraphics[width=0.5\textwidth, height=0.3\textwidth]{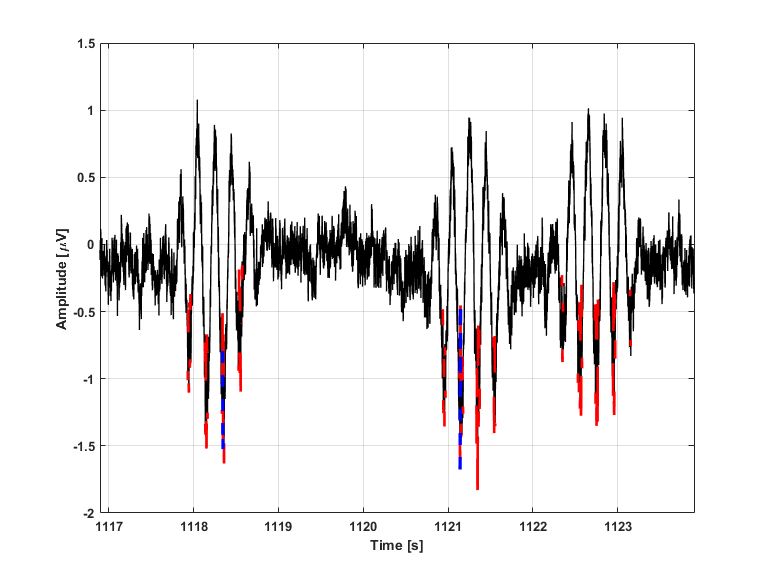}
\end{center}
    \caption{An example of detection of coupling mechanisms in a synthetic EEG signal (black solid line), using REPAC (red line) and the baseline method (blue line).} \label{fig:raw_LFO_HFO} 
\end{figure}
%
%
%
%
%
%
%
%
%
%
We considered the $SNR$ values in $\{-18, -10, -5, 0\}$~dB. For each of them, we generated $10$ PAC-like synthetic signals (see Section~\ref{sec:synthesis}), with random values for $f_L$, $f_H$, $m$ and $L$, selected within the following realistic ranges, respectively: $f_L \in \{4, 5, 6, 7, 8\}$~Hz, $f_H \in \{80, 90, 100, 110, 120, 130, 140\}$~Hz, $m   \in \{0.1, 0.3, 0.5, 0.9\}$, $L   \in \{1.5, 3, 5\}$~s.
%
%
%
%
%
%
%
%
For all signals, the LFO band was initially set to $(1, 30)$~Hz for REPAC and to $(1, 15)$~Hz for the baseline method, while the HFO band was initially set to $(60, 150)$~Hz for both methods.
%
%
%
%
%
%
%
%
%
%
In order to evalute the performance of REPAC, a positive sample is defined as a sample where coupling is occurring. Conversely, a negative sample is a sample where no coupling is present.
Thus, the groundtruth is defined as a binary signal where positive samples are set to $1$ and negative samples are set to $0$.
Performance are computed, for each synthetic signal, in terms of sensitivity or true positive rate (TPR), specificity or true negative rate (TNR), and accuracy~\cite{ICC2020}.
%
%
%
%
%
%
%
%
%
%
%
%
%
%
%
%
%
Fig.~\ref{fig:raw_LFO_HFO} shows the application of the two algorithms on a representative $20$~s segment extracted from the same signal of Fig.~\ref{fig:SYNTHSIGN}. 
In this particular case, \gls{repac} selects the refined LFO range as $(3.1, 9.7)$~Hz, HFO as $(43.1, 122.9)$~Hz, $\hat{f}_L = 5.7$~Hz and $\hat{f}_H = 75.6$~Hz, while the baseline method found LFO as $(4.5, 6.1)$~Hz, HFO as $(60, 150)$~Hz (and no further information is given on the estimation of $f_L$ and $f_H$). 
\gls{repac} reaches a level of accuracy ($99.49\%$) similar to the baseline method ($99.61\%$). However, the former outperformed the baseline method in sensitivity ($TPR = 65.21\%$ with \gls{repac}, $TPR = 20.11\%$ with the baseline method), with a similar level of specificity ($TNR = 99.65\%$ with \gls{repac}, $TNR = 100\%$ with the baseline method).
%
%
%
%
%
%
\input{table_performance.tex}
Tab.~\ref{tab:performance} reports the average classification performance across the $10$ PAC-like signals. As it can be seen, \gls{repac} is similarily outperforming the baseline method for other $SNR$ values, even lower.
%
%
%
%
%
%
\begin{figure*}[htpb]
    \begin{center}
	\includegraphics[width=0.45\textwidth]{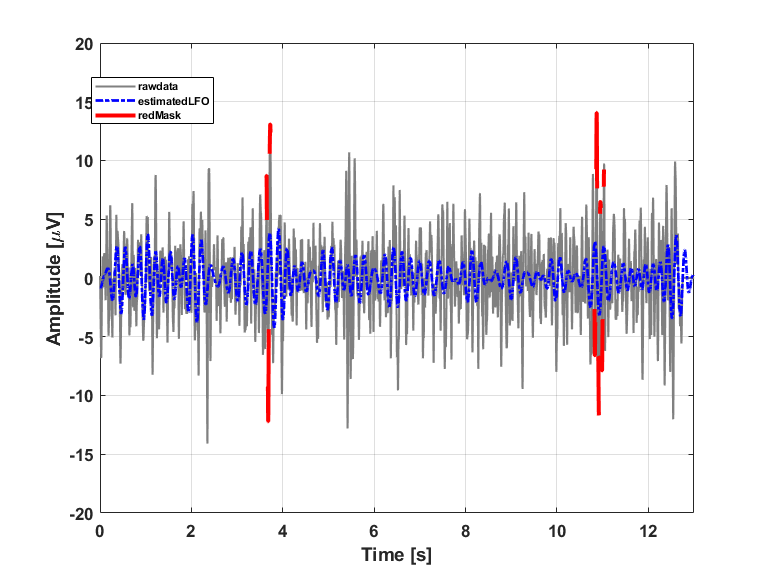}(a) %
   	\includegraphics[width=0.45\textwidth]{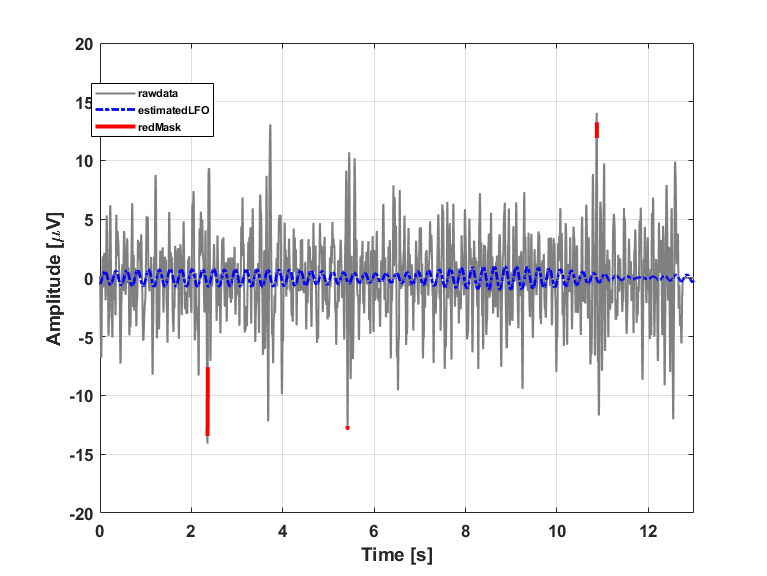}(b)		
    \caption{An example of detection of PAC in a real EEG signal, using (a) REPAC, and (b) the baseline method. In each panel are displayed the raw EEG data (grey solide line), the estimated LFO component (blue line) and the estimated HFO component (red line) for each method, respectively. }\label{fig:Test_real_signal}
    \end{center}
\end{figure*}
Finally, Fig.~\ref{fig:Test_real_signal} reports the preliminary outcome from testing REPAC, in comparison with the baseline method, on a real EEG segment from a healthy subject.
%
%
%
%






%
%
%
%
%

%% file: table_performance.tex
\begin{table*}[htb]
\centering
\caption{Performance comparison between \gls{repac} and the baseline method.} \label{tab:performance}
\begin{tabular}{lccccccc}
\hline
\textbf{SNR [dB]} & \multicolumn{2}{c}{\textbf{Accuracy [$\%$]}} & \multicolumn{2}{c}{\textbf{TPR[$\%$]}} & \multicolumn{2}{c}{\textbf{TNR[$\%$]}}     \\ \hline

    & \textbf{\gls{repac}} & \textbf{Baseline}  & \textbf{\gls{repac}} & \textbf{Baseline}  & \textbf{\gls{repac}} & \textbf{Baseline}         \\ \hline
    
$-18$       & 99.47  & 99.43     &  1.72  &  1.27       & 99.94  & 99.91  \\
$-10$       & 99.49  & 99.61     & 65.21  & 20.11       & 99.65  & 100.0  \\
 $-5$       & 99.18  & 99.62     & 82.07  & 20.68       & 99.27  & 100.0  \\
  $0$       & 99.09  & 99.62     & 84.66  & 20.68       & 99.16  & 100.0  \\ \hline
\end{tabular}
\end{table*}
%
%
%

%% file: Discussion.tex

\gls{repac} allows extracting more precise information about PAC compared to the baseline method and, particularly, more precise estimates of $f_L$, $f_H$, 
the LFO component and the HFO bursts (as seen in Fig.~\ref{fig:raw_LFO_HFO} and in Tab.~\ref{tab:performance}).
As mentioned in section~\ref{sec:Methods}, the SNR, the event duration $L$ and the modulation index $m$ are the three most critical parameters for characterizing PAC~\cite{Makoto2013, Onslow2011, Aru2015, Samiee2017}.
The SNR is typically very low in EEG, 
e.g., $-10$~dB. However, the most previous literature usually set the SNR to much higher values, e.g., above $0$~dB, or it does not even mention it. Only a few studies have reported $SNR$s of $-3$~dB~\cite{Makoto2013}, $-5$~dB~\cite{Samiee2017}, and $-10$~dB~\cite{Makoto2013}.
%
%
%
%
%
%
In line with \cite{Aru2015}, we highlight that the frequency content of LFO is not limited to the carrier frequency $f_L$, but its band should include side-peaks due to its AM modulation.
In the frequency domain, the widening of the LFO frequency band is roughly proportional to $\frac{1}{L}$ (depending also on the window shape) with $L$ standing for the window duration.
According to these new considerations, \gls{repac} could outperform the baseline method in the estimation of $\hat{f}_L$.
On the other hand, the HFO band is widely chosen by visual inspection or based on a-priori empirical field knowledge.
%
However, here for the first time (as far as the author knows), the HFO is recognized as a frequency comb~\cite{FemtoSecondLASERS}.
%
%
%
As such, HFO can be properly decoded by \gls{repac}, as described in section~\ref{sec:Methods}.
This is also clear from Fig.~\ref{fig:raw_LFO_HFO} and Fig.~\ref{fig:Test_real_signal}, where \gls{repac} is shown to outperform the baseline method in a PAC-like synthetic signal and in a real one, respectively.
%
%
%
%
%
%







%% file: REPAC_ICASSP21_main.bbl
\begin{thebibliography}{10}

\bibitem{Canolty2010}
R.~T. Canolty and R.~T. Knight,
\newblock ``The functional role of cross-frequency coupling,''
\newblock {\em Trends Cogn. Sci}, vol. 14, pp. 506--515, 2010.

\bibitem{Fontolan2014}
L.~Fontolan et~al.,
\newblock ``The contribution of frequency-specific activity to hierarchical
  information processing in the human auditory cortex,''
\newblock {\em Nat. Comm.}, vol. 5, pp. 4694, 2014.

\bibitem{Healthcom2018}
G.~Cisotto, A.~V. Guglielmi, L.~Badia, and A.~Zanella,
\newblock ``Classification of grasping tasks based on eeg-emg coherence,''
\newblock in {\em 2018 IEEE 20th International Conference on e-Health
  Networking (Healthcom)}, Ostrava (Czech Republic), pp. 1--6.

\bibitem{Tort2010}
A.~B.~L. Tort, R.~Komorowski, H.~Eichenbaum, and N.~Kopell,
\newblock ``Measuring phase-amplitude coupling between neuronal oscillations of
  different frequencies,''
\newblock {\em J. Neurophysiol}, vol. 104, pp. 1195--1210, 2010.

\bibitem{DeHemptinne2013}
C.~De~Hemptinne et~al.,
\newblock ``Exaggerated phase-amplitude coupling in the primary motor cortex in
  parkinson disease,''
\newblock {\em PNAS}, vol. 110, no. 12, pp. 4780--4785, 2013.

\bibitem{Fell2011}
J.~Fell and N.~Axmacher,
\newblock ``The role of phase synchronization in memory processes,''
\newblock {\em Nat. Rev. Neurosci}, vol. 12, pp. 105--118, 2011.

\bibitem{Lachaux1995}
J.~P. Lachaux, E.~Rodriguez, J.~Martinerie, and F.~J. Varela,
\newblock ``Measuring phase synchrony in brain signals,''
\newblock {\em Human Brain Mapping}, vol. 8, no. 4, pp. 194--208, 1999.

\bibitem{Makoto2013}
M.~Miyakoshi et~al.,
\newblock ``Automated detection of cross-frequency coupling in the
  electrocorticogram for clinical inspection,''
\newblock in {\em Proc. IEEE EMBS}, July 2013, pp. 3282--3285.

\bibitem{Ball2009}
T.~Ball et~al.,
\newblock ``Signal quality of simultaneously recorded invasive and non-invasive
  eeg,''
\newblock {\em NeuroImage}, vol. 46, pp. 708--716, 2009.

\bibitem{Globecom2018}
G.~Cisotto, A.~V. Guglielmi, L.~Badia, and A.~Zanella,
\newblock ``Joint compression of eeg and emg signals for wireless biometrics,''
\newblock in {\em 2018 IEEE Global Communications Conference (GLOBECOM)}, Abu
  Dhabi (United Arab Emirates), pp. 1--6, 2018 IEEE Global Communications
  Conference (GLOBECOM).

\bibitem{EEGLAB}
A.~Delorme and S.~Makeig,
\newblock ``Eeglab: an open source toolbox for analysis of single-trial eeg
  dynamics including independent component analysis,''
\newblock {\em J. Neurosci. Met}, vol. 33, no. 3, pp. 9--21, 2004.

\bibitem{Mitra}
S.~K. Mitra,
\newblock {\em Digital signal processing: a computer-based approach},
\newblock McGraw-Hill, 2007.

\bibitem{Schwartz}
M.~Schwartz,
\newblock {\em Information transmission, modulation and noise},
\newblock McGraw-Hill, 1990.

\bibitem{ICC2013}
G.~Cisotto et~al.,
\newblock ``Brain-computer interface in chronic stroke: An application of
  sensorimotor closed-loop and contingent force feedback,''
\newblock in {\em 2013 IEEE International Conference on Communications (ICC)}.
  2013, pp. 4379--4383, Budapest.

\bibitem{Sauseng2016}
B.~Berger et~al.,
\newblock ``Brain oscillatory correlates of altered executive functioning in
  positive and negative symptomatic schizophrenia patients and healthy
  controls,''
\newblock {\em Front. Psychol}, vol. 7, pp. 705, 2016.

\bibitem{FemtoSecondLASERS}
S.~T. Cundiff and J.~Ye,
\newblock ``Colloquium: Femtosecond optical frequency combs,''
\newblock {\em Rev. Mod. Phys}, vol. 75, pp. 325--342, 2003.

\bibitem{ICC2020}
G.~Cisotto, M.~Capuzzo, A.~V. Guglielmi, and A.~Zanella,
\newblock ``Feature selection for gesture recognition in internet-of-things for
  healthcare,''
\newblock in {\em 2020 IEEE International Conference on Communications (ICC)},
  Dublin (Ireland), pp. 1--6.

\bibitem{Onslow2011}
A.~C.~E. Onslow, R.~Bogacz, and M.~W. Jones,
\newblock ``Quantifying phase--amplitude coupling in neuronal network
  oscillations,''
\newblock {\em Progr. Biophys. Mol. Biol}, vol. 105, no. 1, pp. 49--57, March
  2011.

\bibitem{Aru2015}
J.~Aru et~al.,
\newblock ``Untangling cross-frequency coupling in neuroscience,''
\newblock {\em Curr. Op. Neurobiol.}, vol. 31, pp. 51--61, 2015.

\bibitem{Samiee2017}
S.~Samiee and S.~Baillet,
\newblock ``Time-resolved phase-amplitude coupling in neural oscillations,''
\newblock {\em NeuroImage}, vol. 159, pp. 270--279, 2017.

\bibitem{Lisman2013}
J.~E. Lisman and O.~Jensen,
\newblock ``The theta--gamma neural code,''
\newblock {\em Neuron}, vol. 77, pp. 1002--1016, 2013.

\bibitem{Cole2017}
S.~R. Cole and B.~Voytek,
\newblock ``Brain oscillations and the importance of waveform shape,''
\newblock {\em Trends Cogn. Sci}, vol. 21, no. 2, pp. 137--149, February 2017.

\bibitem{Silvoni2014}
S.~Silvoni et~al.,
\newblock ``"kinematic and neurophysiological consequences of an
  assisted-force-feedback brain-machine interface training: a case study",''
\newblock {\em Frontiers in neurology}, vol. 4, pp. 173, 2013.

\bibitem{AESM2018}
G.~Cisotto and S.~Pupolin,
\newblock ``Evolution of ict for the improvement of quality of life,''
\newblock {\em in IEEE Aerospace and Electronic Systems Magazine}, vol. 33, no.
  5-6, pp. 6--12, 2018.

\bibitem{HIT2020}
A.~Zanella, F.~Mason, P.~Pluchino, G.~Cisotto, V.~Orso, and L.~Gamberini,
\newblock ``Internet of things for elderly and fragile people,''
\newblock arXiv preprint, 2020.

\end{thebibliography}
